\documentclass[aps,pre,reprint,superscriptaddress]{revtex4-2}
\usepackage{graphicx}
\usepackage{dcolumn}
\usepackage{bm}
\usepackage{epstopdf}
\usepackage{gensymb}
\usepackage[normalem]{ulem}
\usepackage{amsfonts}
\usepackage{float}
\usepackage{xr}
\usepackage{sidecap}
\usepackage{mathrsfs}
\usepackage{amsmath}
\usepackage{amsthm}
\usepackage{enumerate}
\usepackage{soul}
\usepackage[dvipsnames]{xcolor}
\usepackage{appendix}
\usepackage{physics}
\usepackage{environ}
\usepackage{tikz}
\usetikzlibrary{external}
\usepackage[colorlinks=true, allcolors=blue]{hyperref}
\usepackage{placeins}

\begin{document}

\title{Topological Computation by non-Abelian Braiding in Classical Metamaterials\\
}

\author
{Liyuan Chen$^{1,2}$, Matthew Fuertes$^{3}$, Bolei Deng$^{3*}$\\
\normalsize{$^{1}$Department of Physics, Harvard University,
Cambridge, Massachusetts 02138, USA}\\
\normalsize{$^{2}$Harvard John A. Paulson School of Engineering and Applied Sciences}\\
\normalsize{Harvard University,
Cambridge, Massachusetts 02138, USA}\\
\normalsize{$^{3}$School of Aerospace Engineering, Georgia Institute of Technology Atlanta, Georgia 30332, USA}\\
\vspace{2pt}
\normalsize{$^{*}$Correspondence to: bolei.deng@gatech.edu}\\
}



\begin{abstract}
We propose a realization of the one-dimensional Kitaev topological superconductor in classical mechanical metamaterials. By designing appropriate braiding protocols, we demonstrate that the system's mid-gap vibrational modes, termed classical Majorana zero modes (MZMs), accurately reproduce the braiding statistics predicted by quantum theory. Encoding four MZMs as a classical analog of a qubit, we implement all single-qubit Clifford gates through braiding, enabling the simulation of topological quantum computation in a classical system. Furthermore, we establish the system's topological protection by demonstrating its robustness against mechanical defects. This work provides a novel framework for exploring topological quantum computation using classical metamaterials and offers a pathway to realizing stable vibrational systems protected by topology.
\end{abstract}

\maketitle

\section{Introduction}

Topological quantum computation (TQC) is one of the promising routes toward large-scale quantum computation~\cite{Kitaev2003Annals_FT_anyons,Freedman2002_Fibonacci_UQC,Nayak2008RMP_TQC}, where the information is protected by topology against local noise. Majorana fermions~\cite{1937Majorana}, or Majorana zero modes (MZMs) in the $p$-wave topological superconductor, i.e., Kitaev chain~\cite{Kitaev_2001,Bravyi_2002_Fermionic_QC,Lutchyn_2010,Alicea_2010_Majorana,Alicea2011,Sarma2015}, are considered one of the most promising candidates to realize TQC. Their non-Abelian braiding statistics enable the realization of quantum computation in a fault-tolerant manner. However, although various recent advances indicate the potential observation of MZMs in condensed matter systems~\cite{Wang_Nature_2022,Wang_Exp_2023,Bordin_Exp_2023,Dvir_2023,Zatelli_2024,Bordin_PRL_2024,Ghorashi_2024_PRL,head20243dactivenematicdisclinations,Aghaee2025}, it is still far from completely manipulating these quasiparticles for practical topological quantum computation. 

Classical systems, including phononic and acoustic systems~\cite{Huber_2015_Science,Wang_2015_PRL,Khanikaev2015_NC_acoustic,He_2016_NP,Prodan2017_NC,Miniaci_2018_PRX,Prodan_2018_PRB,Chen_2019_AM,Prodan_2020_PRL,Xue_2022_Nat_Rev_Matter,Liu_2023_PRApplied,Qian_2023_PRR}, photonic crystals~\cite{Raghu_2008_PRA,Wu_2015_PRL_Photonic,Yang_2018_PRL_Photonic,Ozawa_2019_RMP,Wang2023_NC_Photonics,Khanikaev2024_NC_Photonics}, and electronic circuits~\cite{Imhof2018_Electric,Lee_2018_Phys_Comm_Electric,Wang_2020_NC_Electric,Ezawa_2020_PRB_Electric,Dong_2021_PRR,Stegmaier_2024_PRR_Electric} provide an alternative to simulate and understand single-particle behaviors in topological quantum systems. Among these, phononic metamaterials stand out due to their stability and feasibility in engineering, making them ideal candidates for the realization of the Kitaev chain and the study of braiding Majorana zero modes (MZMs)~\cite{Prodan2017_NC,Prodan_2018_PRB,Prodan_2020_PRL,Liu_2023_PRApplied,Qian_2023_PRR}. However, previous studies in this direction have only demonstrated the braiding of classical MZMs from the same pair, lacking a general protocol for braiding arbitrary pairs of MZMs. Consequently, current studies are far from providing faithful classical simulations of topological quantum computation. Moreover, earlier numerical studies have focused primarily on solving the eigenstates of the dynamical matrices rather than performing first-principles simulations of the systems. This discrepancy might lead to a gap between theoretical predictions and actual dynamics. Lastly, the effective manipulation of MZMs requires smooth tuning of the system’s parameters, a capability not straightforwardly achievable with existing strategies.

In this work, we propose a feasible classical metamaterial design for realizing the Kitaev topological superconductor, where coupling constants can be easily tuned to manipulate MZMs. We develop a braiding protocol that allows arbitrary pairs of mid-gap MZMs to be exchanged and, through first-principles simulations, explicitly demonstrate that their braiding obeys the algebra of the braid group~\cite{PhysRevLett.86.268}. By encoding four MZMs into a classical analog of a qubit, we implement all single-qubit Clifford gates via braiding, enabling a faithful simulation of single-qubit topological quantum computation. Furthermore, we establish the system’s robustness against mechanical defects, providing a classical analog of fault-tolerant topological quantum computation. This work facilitates the study of topological quantum computation by classical metamaterials and offers new insights into designing stable classical wave systems with topological protection.

\section{1D mechanical Kitaev chain model} 
\label{sec:1D_Kitaev_Chain}

In this section, we introduce the realization of the $1$-dimensional Kitaev chain using mechanical metamaterials.
The vibration of a periodic mechanical system with multiple degrees of freedom (DOFs) is described by Newton's equations in momentum space~\cite{Hussein_2014_Review_Phonon}: 
\begin{equation} \label{eqn:governing_eqn}
    \left[\omega^2 - D(\textbf{k})\right] U(\textbf{k}) = 0\;,
\end{equation}
where $\omega$ is the frequency and $\textbf{k}$ is the wave-vector in the first Brillouin zone. Here, $U(\textbf{k})=\{u^a_k\}$ represents the displacement of the $a$-th degree of freedom in the unit cell, relative to its equilibrium position. $D(\textbf{k}) = M^{-1}K(\textbf{k})$ denotes the dynamical matrix with $M$ and $K(\textbf{k})$ the mass and stiffness matrices in $k$-space, respectively. 

The governing equation~\eqref{eqn:governing_eqn} exhibits the same form as the time-independent Schr\"{o}dinger equation of single-particle Hamiltonian in the lattice, with $U$ corresponding to the amplitudes~\cite{mahan1990many,girvin2019modern}. One example is the second quantized Bogoliubov-de Gennes (BdG) Hamiltonian $\hat{H}(\textbf{k}) = \Psi^\dagger(\textbf{k})H(\textbf{k})\Psi(\textbf{k})$ in the Nambu basis $\Psi(\textbf{k}) = (c(\textbf{k}),c^\dagger(-\textbf{k}))^T$, where $c^\dagger(\textbf{k})$ and $c(\textbf{k})$ are the creation and annihilation operators of a fermionic mode with momentum $\textbf{k}$, respectively~\cite{de1999superconductivity}. The BdG Hamiltonian describes the low-energy single-particle physics of superconductors, whose quasiparticle excitations with momentum $\textbf{k}$ are described by the eigenstates of the first quantized Hamiltonian $H(\textbf{k})$.

Throughout this paper, we focus on the $1$-dimensional Kitaev $p$-wave topological superconductor which supports Majorana zero modes (MZMs) for topological quantum computation~\cite{Kitaev_2001}.
In the Majorana basis, the first quantized Hamiltonian $H(\textbf{k})$ reads~\cite{Qi_2011,bernevig2013topological}
\begin{equation} \label{eqn:Kitaev_chain_Hamiltonian}
    H(\textbf{k}) = (t\cos(\textbf{k})-\mu)\sigma_2-\Delta_{x}\sin(\textbf{k})\sigma_1 - \Delta_{y}\sin(\textbf{k})\sigma_3
\end{equation}
where $\sigma_1,\sigma_2$ and $\sigma_3$ are the Pauli matrices. In \eqref{eqn:Kitaev_chain_Hamiltonian}, the parameters $\mu,t$ and $\Delta = \Delta_{x}+i\Delta_{y}$ correspond to the chemical potential, the hopping amplitude and the complex-valued superconducting pairing parameter, respectively. In the superconducting phase where $|\Delta|$ is a positive constant, the system is topologically nontrivial when $|\mu|<|t|$ and trivial when $|\mu|>|t|$. The MZMs are trapped at the domain walls of topologically distinct regions~\cite{Kitaev_2001}.

By engineering $D(\textbf{k})$ in \eqref{eqn:governing_eqn}, one can realize the Kitaev chain (i.e., \eqref{eqn:Kitaev_chain_Hamiltonian}) and partially reproduce the computation based on MZMs in classical metamaterials. The $H(\textbf{k})$ in \eqref{eqn:Kitaev_chain_Hamiltonian} corresponds to a purely imaginary Hamiltonian $H$ in real space. 
By introducing an auxiliary subspace and tensoring $H$ with the imaginary Pauli matrix $\tau_2$, we obtain a purely real dynamical matrix $D$ as:
\begin{equation} \label{eqn:real_space_dynamical_matrix}
    D = \tau_2 \otimes H + \omega_{0}^2 I\;,
\end{equation}
where $\omega_{0}$ is the natural frequency of the system. The detailed matrix $D$ in \eqref{eqn:real_space_dynamical_matrix} can be found in Appendix~\ref{appendix:details_dyn_matrix}. We use $U = \{u^a_i\}$ to denote the small displacement of the $a$-th degree of freedom at the $i$-th unit cell.\\

 \begin{figure}[t]
    \begin{center}
    \includegraphics[width=0.87\columnwidth]{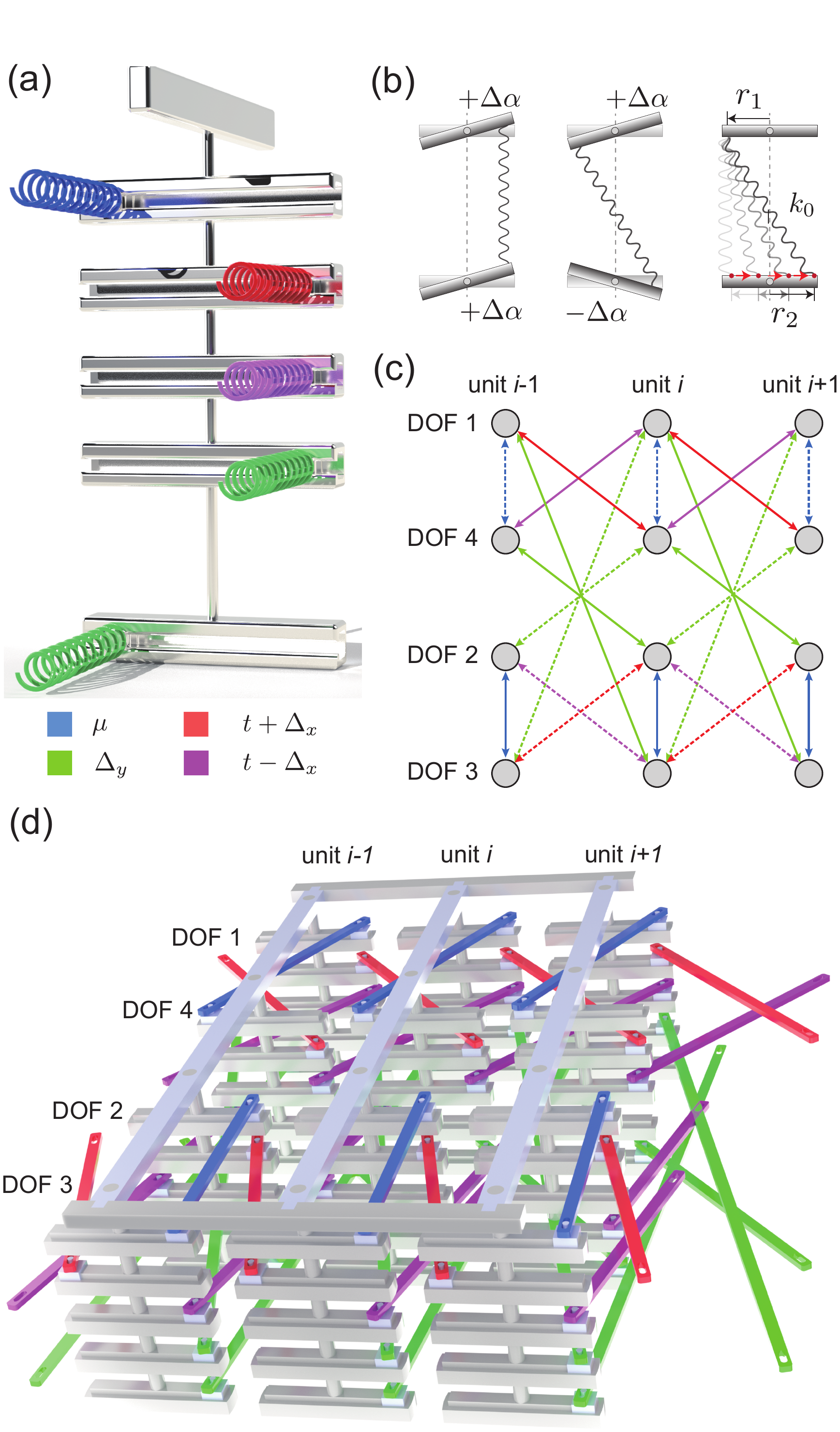} 
        \caption{\textbf{Design Principles of the mechanical metamaterial system that support Majorana Zero Modes (MZMs)} (a) Each degree of freedom (DOF) of the metamaterial features a hanging spindle that can rotate from its equilibrium position along the perpendicular axis, with couplings between various DOFs realized by springs whose ends are movable along horizontal tracks, matching the colors used in the dynamical matrix (see Appendix~\ref{appendix:details_dyn_matrix}). (b) The tuning of the effective spring constants is viewed from above the spindle, showing that the coupling constant between two DOFs can be positive or negative depending on whether the spring ends are on the same side or on opposite sides, with $\Delta\alpha$ representing the rotation angle of the spindle; the coupling constant is adjusted by moving the ends of the spring. (c) The spring connections conform to the configurations in the dynamical matrix $D$, where solid and dashed lines take corresponding and opposite values to their colors, as shown previously. (d) A small segment of the final assembled metamaterials, featuring three consecutive units labeled $i-1$, $i$, and $i+1$, each with four spindles representing four DOFs.} 
        \label{fig1}
        \vspace{-15pt}
    \end{center}
\end{figure}

We design a mechanical metamaterial system, which precisely embodies the dynamical matrix $D$ detailed in \eqref{eqn:real_space_dynamical_matrix}, as depicted in Fig.~\ref{fig1}. This system comprises $N$ units, each featuring 4 degrees of freedom (DOFs). Each DOF is implemented through the rotation of a spindle, as shown in Fig.~\ref{fig1}(a). The rotation angle of the $a$-th spindle in the $i$-th unit is denoted by $u^a_i$. To achieve the desired couplings, each spindle is equipped with rigid bars that are interconnected with springs to neighboring spindles. The connection points on these bars are adjustable along a track, enabling the adiabatic tuning of the coupling strengths. The coupling strength, represented by the matrix entries $D_{ij}$, quantifies the interaction between the $i$-th and $j$-th DOFs. Specifically, such coupling strength is release through springs and rigid bars as $k_0 r_1r_2/J$, where $r_1$ and $r_2$ are the positions of attachment on the bars (refer to the right panel of Fig.~\ref{fig1}(b)), $k_0$ is the spring constant, and $J$ is the rotational inertia of the spindle. Notably, the sign of the coupling (positive or negative) depends on the attachment positions $r_1$ and $r_2$ of the springs, as illustrated in Fig.~\ref{fig1}(b). 

\begin{figure*}[t]
    \begin{center} \includegraphics[width=1.9
    \columnwidth]{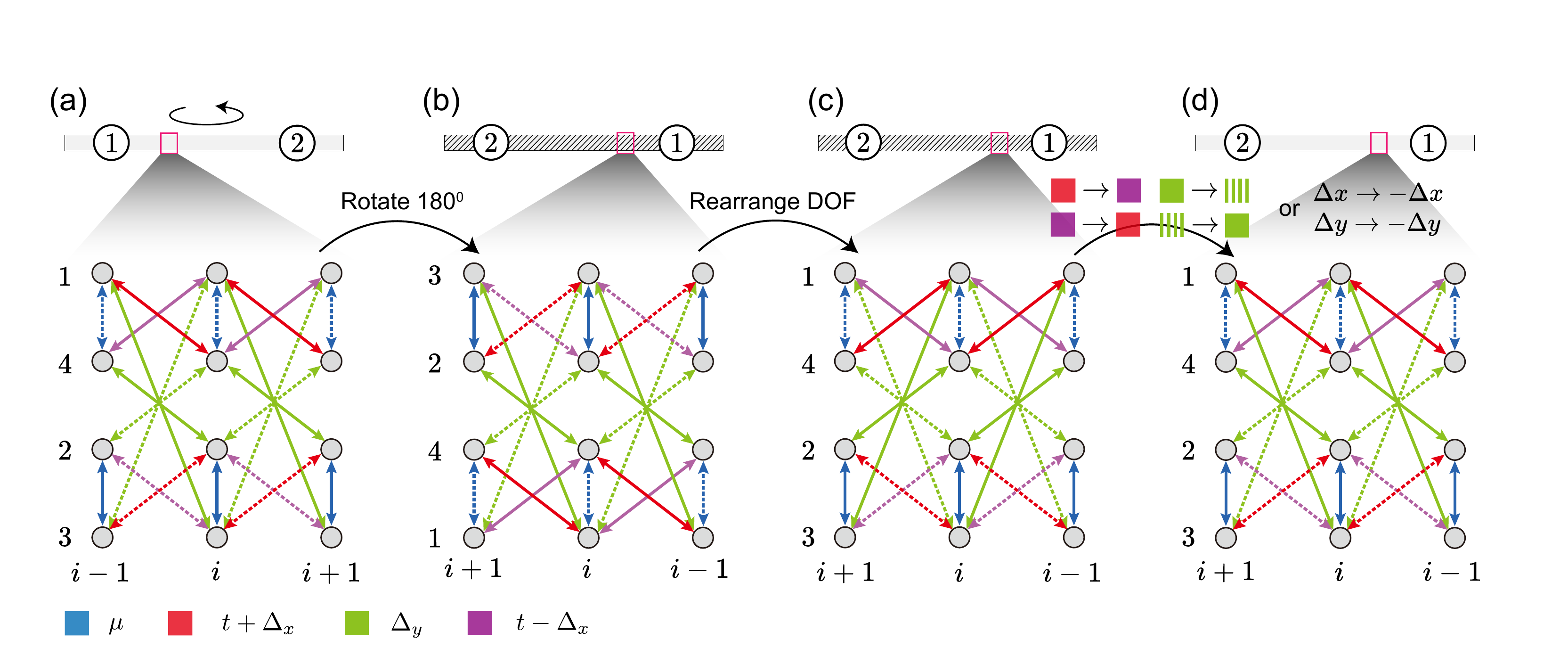}
        \caption{\textbf{Braiding of Majorana Zero Modes (MZMs) in our classical metamaterial system.} (a) The initial configuration of a sector with a pair of MZMs. (b) The entire sector is flipped by rotating it $180^\circ$. (c) Rearrangement of DOFs to match the original configuration. (d) Spring connections are further tuned to restore the original setup, specifically transforming red $(t+\Delta x)$ to purple $(t-\Delta x)$ and green $\Delta_y$ to dashed green $-\Delta_y$, resulting in the identical classical system with flipped MZMs.}
        \label{fig_flip}
        \vspace{-15pt}
    \end{center}
\end{figure*}

The specific spring connections between the DOFs of the units are displayed in Fig.~\ref{fig1}(c), with colors indicating different terms in the matrix $D$. Solid and dashed lines distinguish between positive and negative coupling strengths, respectively. Finally, the entire assembly of spindles and springs constitutes the real mechanical metamaterial are demonstrated in Fig.\ref{fig1}(d). This metamaterial consists of $N$ columns, each representing one unit with four spindles functioning as a single DOF. 
In the remainder of this article, we will simulate the dynamical behavior of this metamaterial system and explore non-Abelian braiding by moving the attachment points of the springs on the rigid bars.

\section{Topological operations}
\label{sec:topo_operations}

Dynamically, when exciting the classical Kitaev chain in an appropriate frequency, there are end states located at the interfaces between topologically trivial and non-trivial regions. We denote these states by classical MZMs, and MZMs in short.
We expect the classical MZMs to be capable of topological computation as their quantum counterparts.
Braiding, defined as the counterclockwise exchange of the positions of anyons, constitutes a fundamental operation in topological quantum computation~\cite{Kitaev_2006}.
In this section, we demonstrate that our mechanical system faithfully simulates the braiding statistics of MZMs predicted by quantum theory. Based on that, we verify that the braiding of classical MZMs satisfies the Yang-Baxter equations, enabling a classical representation of the braid group.

MZMs are always created in pairs at both ends of topologically non-trivial regions described by Kitaev (2001)~\cite{Kitaev_2001}. Within such sector, braiding MZMs is relatively straightforward using the common T-junction method, discussed in detail by Alicea et al. (2010)~\cite{Alicea_2010_Majorana}, which we will further explore in Section~\ref{sec:intra_sector_braiding}. However, this method is not
directly applicable to classical systems. To address this limitation, we have developed new protocols that differ from the traditional T-junction approach. These protocols are specifically designed to facilitate inter-sectional braiding of MZMs, as detailed in Section~\ref{sec:inter_sector_braiding}.

The main idea of our protocol is inspired by the physics of the one-dimensional quantum Kitaev chain. In such a system, the complex-valued order parameter $\Delta$ has a principal angle $\theta = \mathrm{arg}(\Delta_x + i\Delta_y)$, as detailed by De Gennes (1999)~\cite{de1999superconductivity}. Braiding two MZMs involves exchanging their positions, effectively enacting a rotation that locally transforms $\theta$ from $0$ to $\pi$. Critically, this transformation is local to the two MZMs, keeping the actual physical parameter $\theta$ invariant within the system. In our mechanical system, this process can be accurately simulated as illustrated in Fig.~\ref{fig_flip}. The figure contains four subfigures, each depicting one of the four steps of the simulation. In both the upper and lower halves of each subfigure, we present a schematic (a band) of the metamaterial alongside a representative segment showing the spring connections. The flipping of the chain is indicated by shading within the band, while the variations in the spring connections are highlighted through different colors and shapes of the spring lines.


To perform braiding in our classical system, we begin with a sector containing a pair of MZMs, as depicted in Fig.~\ref{fig_flip}(a). The braiding process starts by flipping the entire sector, which results in a new chain rotated by $180^\circ$ from the original configuration (see Fig.~\ref{fig_flip}(b)). Although this action swaps the positions of the two MZMs, the coupling springs in the chain differ from those in the original setup. To restore the chain to its original configuration, we first rearrange the DOFs to reflect the initial arrangement (see Fig.~\ref{fig_flip}(c)). Comparing Fig.~\ref{fig_flip}(c) to Fig.~\ref{fig_flip}(a), it is evident that the spring arrangements are altered; specifically, the red and purple springs are switched, and the green spring couplings are transformed into their opposite values. To revert the current structure to the original setup, we then gradually flip these switched springs back, specifically transforming red ($t+\Delta x$) to purple ($t-\Delta x$) and green $\Delta_y$ to dashed green $-\Delta_y$. This process corresponds to a $\Delta_x \to -\Delta_x$ and $\Delta_y \to -\Delta_y$ transformation, serving as a classical analog to the local $\theta = 0 \to \pi$ shift experienced by the two MZMs. This final adjustment is illustrated in the fourth step of the process, ensuring that the chain fully returns to its initial state, effectively completing the classical simulation of MZM braiding.

A fundamental requirement of our braiding protocol is that it preserves the system’s topology, ensuring the protection of the logical information encoded in MZMs. As reviewed in Appendix~\ref{appendix:review_Kitaev_chain}, the one-dimensional quantum Kitaev chain possesses a spectral energy gap, with MZMs appearing as mid-gap modes localized at zero energy under open boundary conditions. As long as this gap remains open—preventing a topological phase transition—MZMs remain isolated from bulk modes, thereby preserving the encoded logical information. This spectral gap serves as the foundation of topological protection. In our classical Kitaev chain, domain walls act as effective open boundaries, leading to mid-gap vibrational modes in the frequency spectrum. To maintain topological protection, it is essential that braiding operations do not close the spectral gap. We verify that this condition holds in Appendix~\ref{appendix:review_Kitaev_chain}.



In the subsequent sections, we will explicitly demonstrate the application of our protocol for braiding arbitrary pairs of Majorana zero modes (MZMs). Specifically, we simulate two sectors, each containing a pair of MZMs distributed across 40 units, resulting in a dynamic simulation involving $N=80$ units in total. The middle and boundary regions of each sector are set to be topologically non-trivial ($\mu = 0$) and trivial ($\mu = 2$), respectively. This configuration ensures that each sector harbors a pair of MZMs at the domain walls defined by $\mu$. To simulate the dynamic vibrations of the system, we employ the Runge–Kutta method, utilizing the MATLAB function \texttt{ode45}. The parameters for our simulation are set as follows: $t = 1$, $\omega_0 = 2$, and $|\Delta| = \sqrt{\Delta_x^2 + \Delta_y^2} = 1$. Since the dynamical matrix $D$ (see \eqref{eqn:real_space_dynamical_matrix}) has an intrinsic symmetry $[\tau_2,D] = 0$, the DOFs $2$ and $3$ have the same dynamics as DOFs $1$ and $4$, so it is adequate to read off all information from the latter. Throughout the simulations, we restrain our attention to DOFs 1 and 4, as these are sufficient to reflect the complete dynamics of the entire system.


\subsection{Intra-sector braiding}
\label{sec:intra_sector_braiding}

\begin{figure}[t]
   \begin{center}
   \includegraphics[angle = 0, width=0.9\columnwidth]{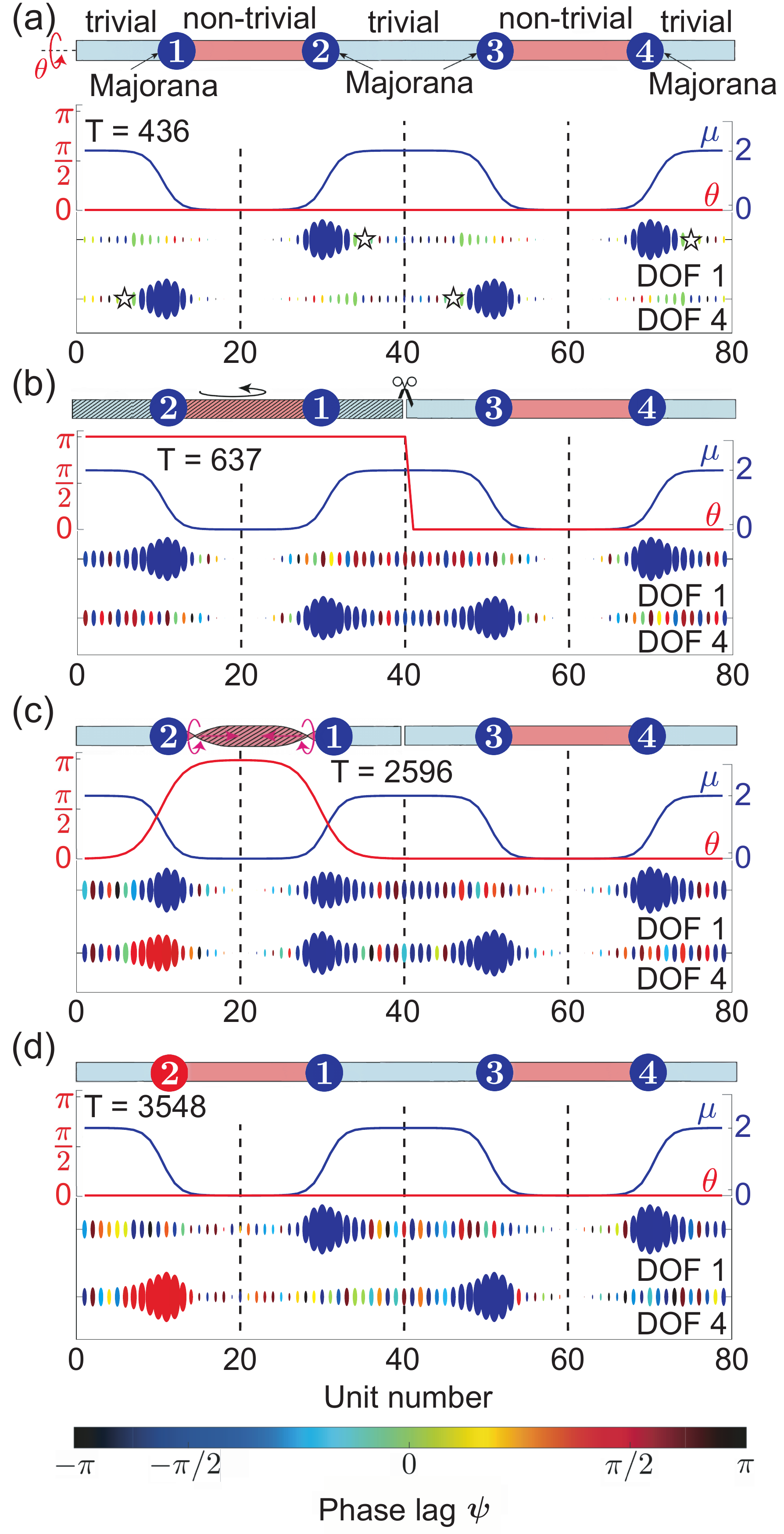}
       \caption{\textbf{Intra-sector braiding of classical Majorana Zero Modes (MZMs).} The braiding process is illustrated in four steps, with each step depicting the system's band schematic and a snapshot of the vibrational modes. (a) {Step 1:} Four MZMs are excited within the system. (b) {Step 2:} The first sector is cut at the topologically trivial regions and flipped. (c) {Step 3:} The parameter $\theta$ is smoothly tuned from $\pi$ to $0$. (d) {Step 4:} Two MZMs are exchanged, resulting in a non-trivial phase $\Delta \psi_2 = \pi$.}
       \label{fig_U12}
       \vspace{-15pt}
   \end{center}
\end{figure}


In this section, we implement our braiding protocol, as illustrated in Fig.~\ref{fig_flip}, to achieve intra-sector braiding of MZMs. The process is documented through chronological snapshots in Fig.~\ref{fig_U12}, where $T$ denotes the non-dimensional simulation time (see Movie S1 for more details). To clearly demonstrate the braiding process, each snapshot depicts, from top to bottom: (i) a band schematic showing the positions of Majorana and the current operation; (ii) the current parameter settings of the dynamic simulation, including the distribution of $\mu$ (blue) and $\theta$ (red); and (iii) the simulation results for all units on the chain with their vibration amplitude indicated by the size of the ellipse and phase difference represented by color. Specifically, the band is colored to indicate topologically non-trivial ($\mu = 0$) and trivial ($\mu = 2$) regions respectively, and the region with $\theta = \pi$ is highlighted by shading. 

We begin by initializing the system, harmonically exciting the vibration at the DOFs marked by stars. The excitation frequency is set to match the natural frequency of the edge mode of the system $\omega_0$---triggering vibrations clustered around the four boundaries between trivial and non-trivial parts and leading to the activation of the two pairs of MZMs in our system. These four MZMs, represented by the Majorana operators $\gamma_1$, $\gamma_2$, $\gamma_3$ and $\gamma_4$, are initially excited to have a phase lag of $\psi_1 = \psi_2 = -\pi/2$ relative to the reference, as indicated by the blue color in Fig.~\ref{fig_U12}(a). Following the protocol outlined in Fig.~\ref{fig_flip}, we perform the braiding by cutting the two edges of the first sector—located in the topologically trivial segments—and flipping the entire sector. As discussed in the previous section, this flipping operation introduces a global shift $\theta = 0 \to \pi$ (see Fig.~\ref{fig_U12}(b)). We then correct this shift by reconnecting the sector to the chain and changing $\theta$ from $\pi$ back to $0$. This adjustment is achieved by smoothly wrapping $\theta$ from both ends of the sector gradually back to $0$ (see Fig.~\ref{fig_U12}(c)). Finally, this series of operations results in the exchange of positions between $\gamma_1$ and $\gamma_2$, introduces a phase shift of $\Delta\psi_2 = \pi$, and completes the intra-sector braiding, as shown in Fig.~\ref{fig_U12}(d).

There are several important facts about the braiding protocol. First, cutting the edges at the topologically trivial regions is permissible because this operation does not alter the system's topology. However, generalizing the protocol to inter-sector braiding—braiding MZMs from different pairs—is not straightforward. This adaptation requires cuts at topologically non-trivial regions, thereby changing the system's topology. A specific protocol for inter-sector braiding will be developed in the following Section~\ref{sec:inter_sector_braiding}.

Second, the tuning of $\theta$ is achieved by smoothly flipping the signs of $\Delta_x$ and $\Delta_y$ while maintaining $|\Delta| = 1$, ensuring the process is adiabatic and topology-preserving. We have verified that the spectral gap is not closed during intra-sector braiding in Appendix~\ref{appendix:review_Kitaev_chain}. The specific manner of changing $\theta$ does not influence the braiding behavior, as long as $\theta$ eventually returns to $0$.

Lastly, the phase shift $\Delta\psi_2 = \pi$ occurs because the braiding introduces a non-trivial Berry phase on the second MZM~\cite{PhysRevLett.86.268}. Thus, our protocol faithfully reproduces the braiding statistics of the first pair of MZMs as: \begin{equation} \gamma_1 \mapsto -\gamma_2,\quad \gamma_2 \mapsto \gamma_1;, \end{equation} up to an overall phase.

\subsection{Inter-sector braiding}\label{sec:inter_sector_braiding}

\begin{figure}[t]
   \begin{center}
   \includegraphics[angle = 0,width=0.9\columnwidth]{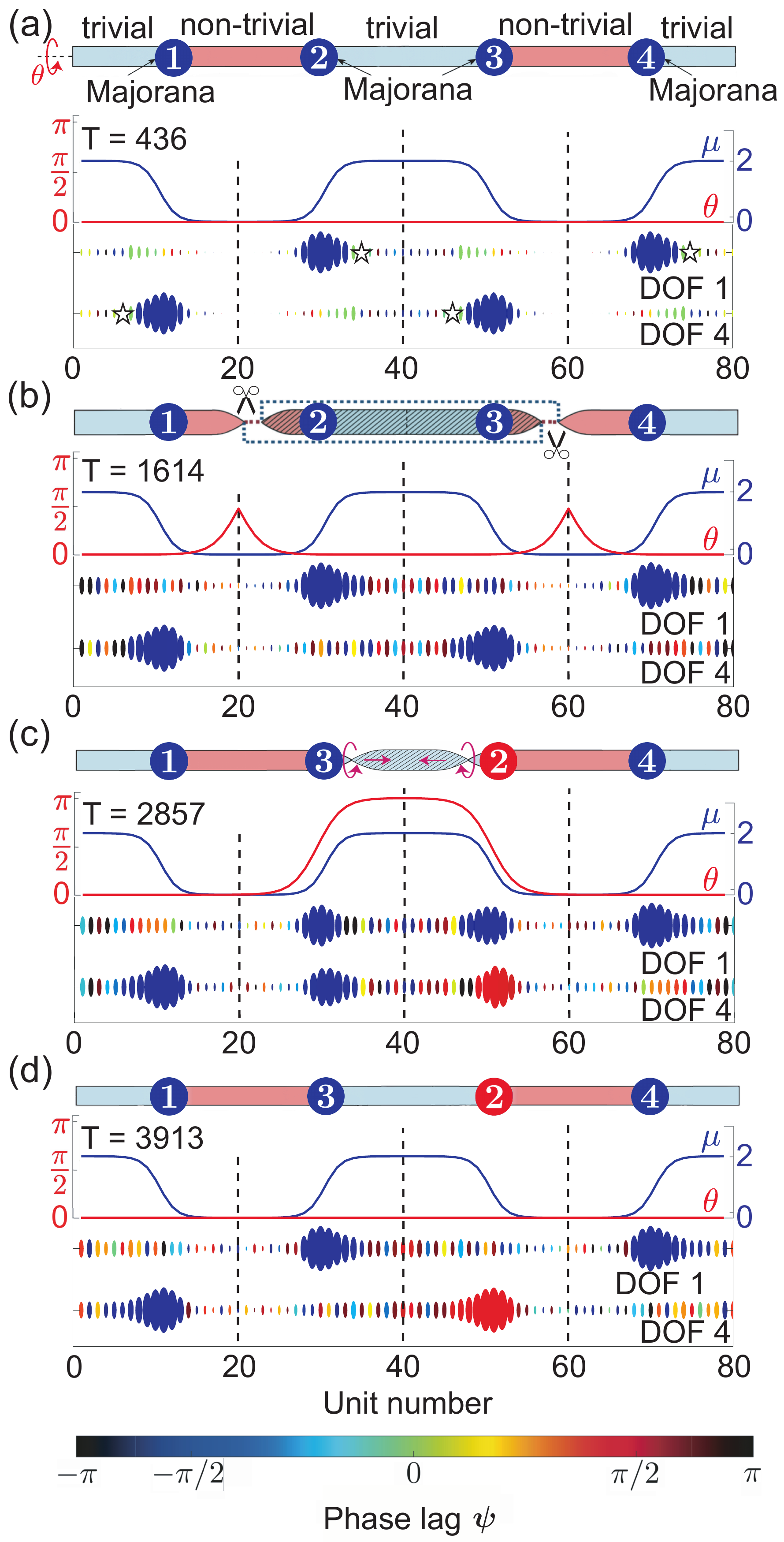}
       \caption{\textbf{Inter-sector braiding of classical Majorana Zero Modes (MZMs).} The process is described in four steps. (a) {Step 1:} Four MZMs are excited within the system. (b) {Step 2:} The segment to be flipped is connected to the target neighbors before its connections with the original neighbors are removed. (c) {Step 3:} The segment is flipped, and the parameter $\theta$ is smoothly tuned from $\pi$ to $0$. (d) {Step 4:} Two MZMs are exchanged, resulting in a non-trivial phase $\Delta \psi_2 = \pi$.}
       \label{fig_U23}
       \vspace{-15pt}
   \end{center}
\end{figure}


In this section, we present our approach for inter-sector braiding of MZMs. It is crucial to note that directly applying the protocol developed in the previous section or as depicted in Fig~\ref{fig_flip} may disrupt computational integrity because these operations involve cutting a segment of the topologically non-trivial part. To address this, we propose a topology-preserving strategy specifically for inter-sector braiding and demonstrate it by braiding $\gamma_2$ and $\gamma_3$, which initially belong to different MZM pairs (see Fig.~\ref{fig_U23}, Movie S2). This inter-sector braiding protocol includes steps similar to those of the intra-sector protocol, with the primary distinction in the flipping step.

As shown in Fig.~\ref{fig_U23}(b), before cutting the segment boundaries (indicated by red dashed lines), we pre-establish new connections with their new neighbors after flipping (shown by blue dashed lines). Additionally, to ensure a smooth transition in the shift of $\theta$ caused by the flipping, we fine-tune the values of $\theta$ at the segment boundaries prior to executing the flip. Following these preparatory steps, the segment can be safely flipped, as each connection within the topologically non-trivial regions acts as a topology-preserving perturbation. Adding or removing such connections does not change the system's topology, provided that at least one connection remains intact. As shown in Fig.~\ref{fig_U23}(c), after the flip, we gradually adjust $\theta$ from $\pi$ back to $0$. This adjustment results in the positional exchange of $\gamma_2$ and $\gamma_3$, accompanied by a phase shift of $\Delta \psi_2 = \pi$ on the second MZM. This careful handling ensures the integrity of the braiding operation while maintaining the topological protection essential for robust computation.


Comparing the configurations of the first and final steps, we can conclude that the two MZMs transform according to the braid group algebra 
\begin{equation}
    \gamma_2 \mapsto \gamma_3 \quad \gamma_3 \mapsto -\gamma_2\;.
\end{equation}
The frequency spectrum of this braiding process, as shown in Appendix~\ref{appendix:review_Kitaev_chain}, implies that this braiding is adiabatic and topology-preserving.

In summary, if we use $U_{i,j}$ to denote the exchange operation of the $i$-th and $j$-th MZMs, we have verified that it realizes the following transformation:
\begin{equation} \label{eqn:braiding_statistics}
    U_{i,i+1}=\left\{ 
    \begin{array}{lr}
         \gamma_{i} \mapsto \gamma_{i+1}  \\
         \gamma_{i+1} \mapsto -\gamma_{i} \\
         \gamma_{j} \mapsto \gamma_{j} \quad \mathrm{for}\  j \neq i\  \mathrm{and}\  j\neq i+1 
    \end{array}
    \right.\;,
\end{equation}
up to a global phase. In Appendix~\ref{appendix:Yang_Baxter_Eqns} and in Movie S3, we also verify that these transformations satisfy the Yang-Baxter equations:
\begin{align} \label{eqn:YB_eqns}
    U_{i,i+1} U_{j,j+1} &= U_{j,j+1} U_{i,i+1}, \quad |i-j|>1\;, \notag\\
    U_{i,i+1} U_{j,j+1} U_{i,i+1} &= U_{j,j+1} U_{i,i+1} U_{j,j+1}, \quad |i-j| = 1\;.
\end{align}
Therefore, with $n$ classical MZMs, we have established a classical representation of the Artin braid group on $n$ strands~\cite{Artin_1947_Braid_Group}. This enables the classical realization of all braiding gates within the one-dimensional Kitaev chain.

\section{Robust computation by Majorana zero models}

The topological properties of classical MZMs enable the implementation of fault-tolerant computation in our classical metamaterial.
By explicitly encoding four MZMs into a qubit, we illustrate that braiding classical MZMs can realize the single-qubit Clifford gates in quantum computation, as detailed in Section~\ref{sec:single_qubit_Clifford}. Section~\ref{sec:FT_braiding} further illustrates that these operations exhibit resilience to noise, analogous to the fault-tolerant capabilities of topological quantum computation. Finally, we provide necessary remarks on realizing universal quantum computation in the classical metamaterial in Section~\ref{sec:UQC}.

\subsection{Single-qubit Clifford gates}
\label{sec:single_qubit_Clifford}

\begin{figure}[t]
    \begin{center}
        \includegraphics[width=0.9\columnwidth]{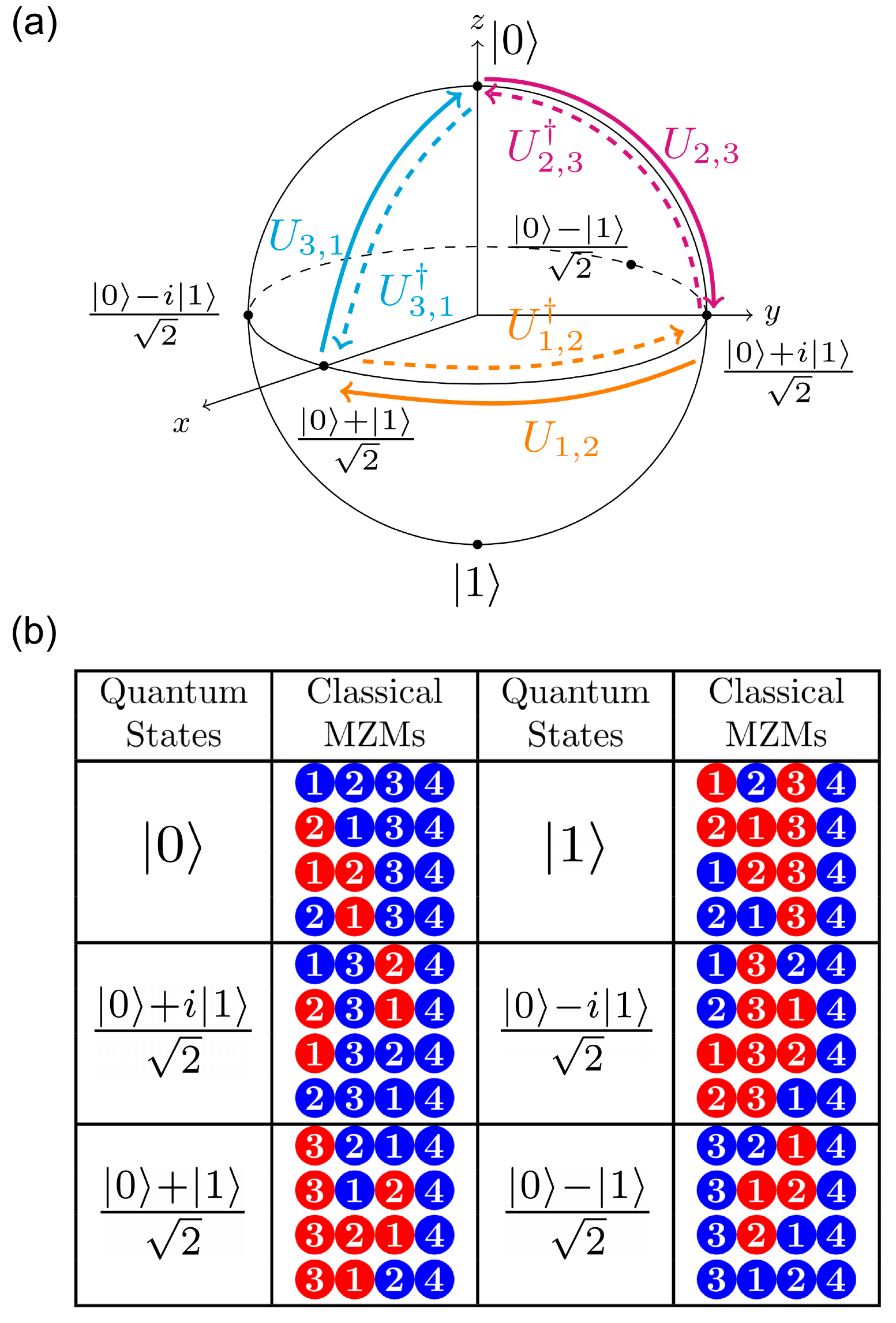}
        \caption{\textbf{The Encoding of Classical MZMs to a Qubit.} (a) The Bloch sphere representation of a qubit, where the exchange operations $U_{i,j}$ correspond to $\pi/2$ rotations along specific axes---for example, $U_{1,2}$ rotates along the $z$ axis, $U_{2,3}$ along the $x$ axis, and $U_{1,3}$ along the $y$ axis. (b) The mapping between the configurations of classical MZMs and quantum states, up to a phase factor.}
        \label{fig_Bloch_Sphere}
        \vspace{-15pt}
    \end{center}
\end{figure}

In this section, we explain how to implement single-qubit Clifford gates by braiding classical MZMs. To achieve this goal, we must define the classical analog of a qubit (quantum bit) and investigate the operations realizable by braiding. Since our encoding is motivated by quantum theory, we start by briefly reviewing quantum computation by MZMs, as proposed by Kitaev and many other authors~\cite{Bravyi_2002_Fermionic_QC,PhysRevLett.86.268,Nayak2008RMP_TQC,Sato_2016_Majorana,Sarma2015}.

One can encode four (quantum) MZMs into one qubit, a 2-dimensional linear space $\mathbb{C}^2$, with the basis $\{\ket{0},\ket{1}\}$.
Specifically, given four Majorana operators $\gamma_1,\gamma_2,\gamma_3,\gamma_4$, we can combine them into two complex fermions $\psi_{12} = (\gamma_1+i\gamma_2)/\sqrt{2}$ and $\psi_{34} = (\gamma_3+i\gamma_4)/\sqrt{2}$. 
Then, we obtain a four-dimensional Hilbert space spanned by $\ket{n_{12}n_{34}}$, where $n_{12},n_{34} \in \{0,1\}$ are fermion occupation numbers~\cite{Sarma2015,Sato_2016_Majorana}. 
Since the Hamiltonian in \eqref{eqn:Kitaev_chain_Hamiltonian} conserves fermion parity, we have a degenerate subspace spanned by states $\{\ket{00},\ket{11}\}$ with even parity. 
Therefore, one can define a qubit by the correspondence $\ket{0} \equiv \ket{00}$ and $\ket{1}\equiv \ket{11}$.

By manipulating quantum MZMs, one can apply quantum operations on the qubits. Specifically, braiding of the MZMs $\gamma_i$ and $\gamma_j$ applies a unitary transformation $U_{i,j} = \exp(i\frac{\pi}{4} \gamma_i\gamma_j)$ on these two operators, as indicated by the algebra in \eqref{eqn:braiding_statistics}.
In the subspace spanned by $\{\ket{00},\ket{11}\}$, the exchange operations have the following matrix representation:
\begin{equation}
    U_{1,2} = (I+i\sigma_z)/\sqrt{2},\quad U_{2,3} = (I+i\sigma_x)/\sqrt{2}.
\end{equation}
If we express the qubit as a directed vector ending on the Bloch sphere, these matrices are $\pi/2$ clockwise rotation along the $z$- and $x$-axes~\cite{Nielsen_Chuang_2010}. The $\pi/2$ rotation along the $y$-axis is realized by $U_{3,1} = U_{1,2}^\dagger U_{2,3}U_{1,2} = (I+i\sigma_y)/\sqrt{2}$, where the Hermitian conjugate $U_{i,j}^\dagger$ represents the inverse process of $U_{i,j}$. All these representations are illustrated in Fig.~\ref{fig_Bloch_Sphere}(a) and Movie S4.

A subset of quantum gates known as Clifford gates can be classically simulated, as outlined by the Gottesman-Knill theorem~\cite{Gottesman-Knill_theorem}. Specifically, braiding Majorana zero modes (MZMs) implements single-qubit Clifford gates~\cite{Sarma2015,Sato_2016_Majorana}, making it relevant to investigate whether these gates can be realized by our classical system. Indeed, as depicted in Fig.~\ref{fig_Bloch_Sphere}(a), the braiding gates only map the qubit to a finite set of points on the Bloch sphere—specifically, the intersection points of the $x,y,z$-axes with the sphere. 
Therefore, we can always classically simulate these single-qubit Clifford gates by introducing proper encoding.

Since there is no classical analog of the fermion occupation number states $\ket{n_{12}n_{34}}$, we establish a special four-to-one correspondence between the configurations of classical MZMs and the qubit states. This correspondence is illustrated in Fig~\ref{fig_Bloch_Sphere}(b), where the configurations provide phase and position information about the classical MZMs.
Each circle represents an MZM as depicted in the snapshots of Figs.~\ref{fig_U12} and \ref{fig_U23}. The color of the circle---red or blue---indicates whether the phase lag is $+\pi/2$ or $-\pi/2$, respectively. Additionally, each MZM is marked with a number that identifies its initial position, aiding in tracking their movements and transformations throughout the braiding process.

To explain this encoding, as an example, we consider the first MZM configuration assigned to the $\ket{0}$ quantum state, where:
\begin{equation} \label{eqn:initial_config}
    \ket{0}\  \leftrightarrow\  \raisebox{-5.5pt}{\includegraphics[]{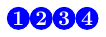}}.
\end{equation}
In this case, the four classical MZMs are in their original positions, each with a $-\pi/2$ phase lag, corresponding to the initial configurations in Figs.~\ref{fig_U12} and \ref{fig_U23}. Braiding $\gamma_1$ and $\gamma_2$ in the MZM configuration results in the following transformation:
\begin{equation}\label{eqn:classical_MZM_U_12}
\begin{gathered}
    \includegraphics[width = 0.22\textwidth]{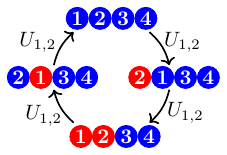}
\end{gathered}
\end{equation}
Meanwhile, in quantum theory, each application of $U_{1,2} = (I+i\sigma_z)/\sqrt{2}$ on $\ket{0}$ introduces a $\pi/4$ phase on the state:
\begin{equation} \label{eqn:quantum_state_U_12}
\begin{gathered}
    \includegraphics[width = 0.4\textwidth]{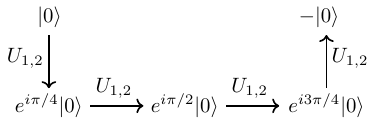}
\end{gathered}
\end{equation}
Thus, by identifying the state $\ket{\psi}$ with $-\ket{\psi}$, the $U_{1,2}$ transforms the classical MZMs and the quantum state $\ket{0}$ in a similar manner. Specifically, four rounds of $U_{1,2}$ application return the states to the initial ones. This fact enables us to identify each intermediate configuration in the loop \eqref{eqn:classical_MZM_U_12} with the quantum state (along with the phase factor) in \eqref{eqn:quantum_state_U_12}.
Moreover, since a global phase is trivial for a quantum state, we further identify these four configurations with $\ket{0}$, as shown in Fig.~\ref{fig_Bloch_Sphere}(b).

Similarly, applying $U_{23}$ on $\ket{0}$ rotates it to the state $(\ket{0}+i\ket{1})/\sqrt{2}$ along the $y$-axis. Then, braiding $\gamma_2$ and $\gamma_3$ transforms the MZM configurations associated with $\ket{0}$ to the ones associated with $(\ket{0}+i\ket{1})/\sqrt{2}$ as follows:
\begin{equation}
    \begin{aligned}
    &\raisebox{-5.5pt}{\includegraphics[]{figure/1234_BBBB.pdf}}\    \xrightarrow{U_{2,3}}\  \raisebox{-5.5pt}{\includegraphics[]{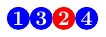}}  \notag\\
     &\raisebox{-5.5pt}{\includegraphics[]{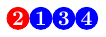}}\  \xrightarrow{U_{2,3}} \ \raisebox{-5.5pt}{\includegraphics[]{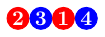}} \;,
\end{aligned}
\end{equation}
with all other transformations obtainable analogously. Furthermore, it is easy to verify that the four configurations associated with $(\ket{0}+i\ket{1})/\sqrt{2}$ are mapped to each other via $U_{3,1}$, as the state is invariant under the rotation along the $y$-axis.

By repeating the above procedures, we can generate all $24$ classical MZM configurations as shown in Fig.~\ref{fig_Bloch_Sphere}(b), and associate them with different quantum states. Indeed, one can find a one-to-one correspondence between the MZM configurations and elements of the permutation group $S_4$.
Since there is a surjection from the braid group $B_4$ to $S_4$, our encoding is justified.

In summary, by properly associating the classical MZM configurations with the quantum states, we can reproduce the single-qubit Clifford operations in our classical system by manipulating classical MZMs. Since the braiding of quantum MZMs is resilient to noise, we expect the computation in our classical system to be robust, as we explore in the following section.

\subsection{Fault tolerance in braiding gates} \label{sec:FT_braiding}
\begin{figure}[t]
    \begin{center}
        \includegraphics[width=1.02\columnwidth]{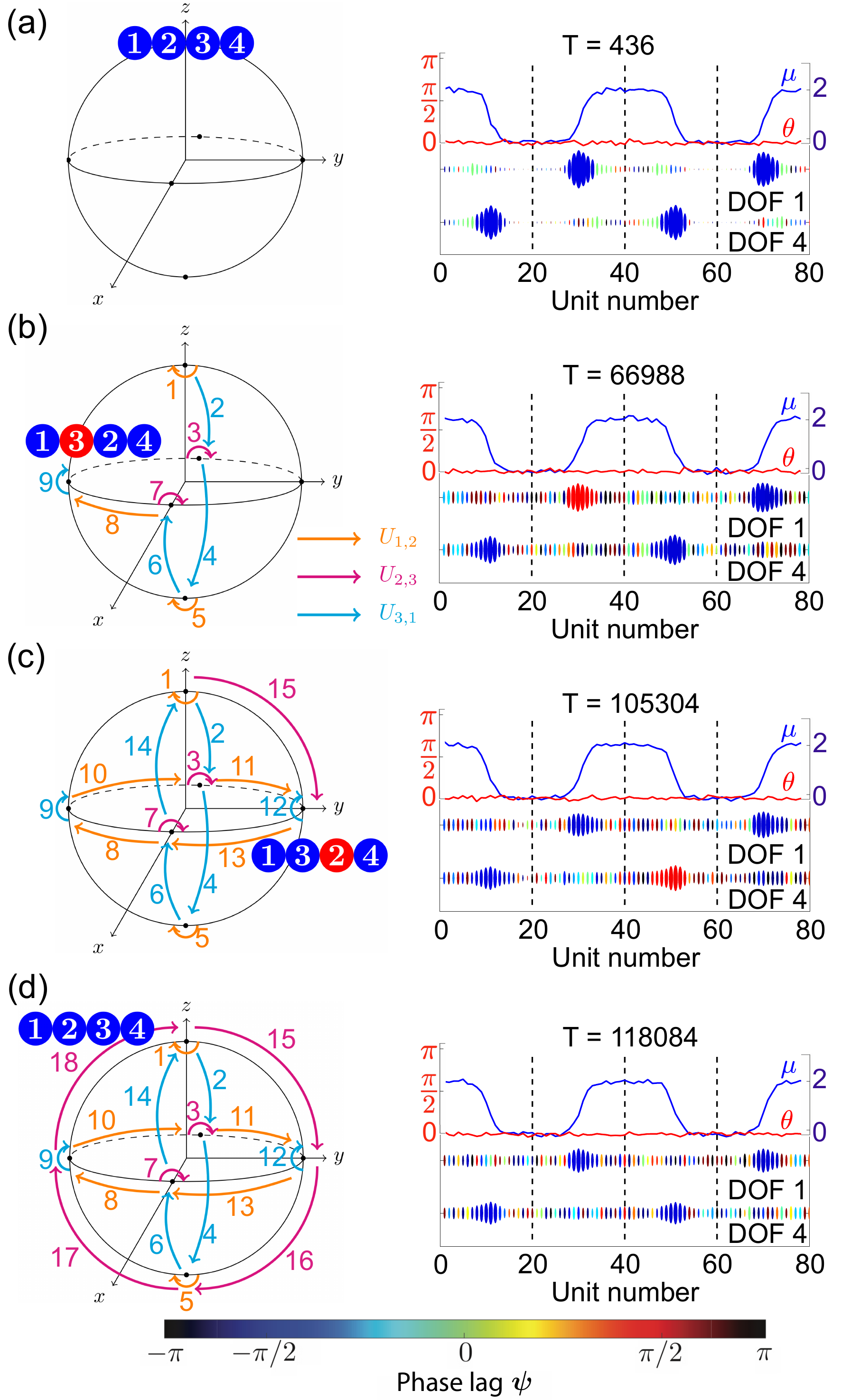}
        \caption{\textbf{Fault tolerance of MZM braiding under noise.} A total of 30 braiding operations are conducted under Gaussian noise perturbations to parameters $\mu,\Delta$. Throughout this generic computation process, the system's state traverses the Bloch sphere, with colored arrows indicating braiding history and labels showing expected MZM configurations. The final configurations remain robust, even in the presence of noise.}
        \label{fig_Fault_Tolerance}
        \vspace{-15pt}
    \end{center}
\end{figure}

The inherent robustness of quantum computation based on MZMs is ensured by topological protection. Specifically, as long as external perturbations do not induce a topological phase transition by closing the spectral gap, their impact on the braiding statistics remains exponentially suppressed~\cite{Sarma2015}. In this section, we demonstrate that an analogous form of topological protection emerges in the classical Kitaev chain.

A key manifestation of this robustness can be observed by simulating a generic computational process using classical MZMs under the influence of noise. When perturbations are sufficiently weak, the computation remains stable over a finite duration before the system undergoes significant degradation. That is, the evolution of the MZM configurations closely follows that of their counterparts in an ideal, noiseless scenario.

This phenomenon is illustrated in Fig.~\ref{fig_Fault_Tolerance} and Movie S5, where we simulate a process in which MZMs are braided, causing the system’s state to traverse the Bloch sphere. The four subfigures depict distinct representative stages of the computation. In each stage, the left panel visualizes the braiding history through indexed colored arrows on the Bloch sphere, alongside the label indicating the expected MZM configuration under idealized noiseless conditions. The right panel presents a snapshot of the system’s instantaneous physical parameters and vibrational behaviors. Before each braiding, we perturb the parameters $\mu\to\mu + \delta \mu$ and $\Delta\to\Delta + \delta \Delta$ of the system by randomly sampling local noise in the Gaussian distribution with variance 0.05. 

The computation begins with an initial MZM configuration, $\ket{0} = \raisebox{-5.5pt}{\includegraphics[]{figure/1234_BBBB.pdf}}$, as illustrated in Fig.~\ref{fig_Fault_Tolerance}(a). A sequence of braiding operations, $U_{3,1}U_{1,2}U_{2,3}U_{3,1}U_{1,2}U_{3,1}U_{2,3}U_{3,1}U_{1,2}$, is then applied to the MZMs, transforming the system into the state $(\ket{0}-i\ket{1})/\sqrt{2} = \raisebox{-5.5pt}{\includegraphics[]{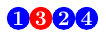}}$ (see Fig.~\ref{fig_Fault_Tolerance}(b)). Further braiding operations $U_{2,3}U_{3,1}U_{1,2}$$U_{3,1}U_{1,2}U_{1,2}$ evolves the system to the state $(\ket{0}+i\ket{1})/\sqrt{2} = \raisebox{-5.5pt}{\includegraphics[]{figure/1324_BBRB.pdf}}$ (see Fig.~\ref{fig_Fault_Tolerance}(c)) before returning to its initial configuration following last operations $U_{2,3}U_{2,3}U_{2,3}$ (see Fig.~\ref{fig_Fault_Tolerance}(d)). 
Throughout these steps, a total of 30 braiding operations are conducted, and our classical MZMs traverse all quantum states on the Bloch sphere. In the presence of perturbations, the classical MZM configurations remain consistent with those in the idealized, noiseless case, demonstrating the fault tolerance of the system.

However, it is important to note that after computation, the MZMs exhibit reduced localization near the domain walls, as shown in Fig.~\ref{fig_Fault_Tolerance}(d). This reduction is a consequence of finite-size effects and perturbations. In a quantum 1D Kitaev chain, MZMs separated by a finite distance experience exponentially small tunneling. Additionally, local perturbations can induce tunneling between MZMs, although this effect remains exponentially suppressed by the spectral gap. Consequently, the MZMs are not perfectly localized after extensive braiding under perturbations. Nonetheless, these effects do not compromise the integrity of the information encoded in the MZMs, thereby ensuring fault tolerance in our simulation. Physically, perturbations in mechanical systems correspond to unavoidable defects, such as variations in spring constants and shifts in connection points. Our results demonstrate that, as long as these mechanical imperfections remain weak, the information encoded in the classical vibrational modes is topologically protected. Furthermore, these modes enable stable computation, highlighting the robustness of the system. Thus, beyond serving as a visualization of topological quantum computation, our system offers a promising framework for realizing resilient classical vibrational systems.

\subsection{Remarks on universal quantum computation} \label{sec:UQC}

In this section, we comment on the limitations of using the classical Kitaev chain for universal quantum computation (UQC). 
Intuitively, since classical systems do not have quantum coherence, it is impossible for them to realize UQC. More specifically, as proposed before, to realize the universal quantum computation by Majorana modes, one needs to implement a universal gate set $\{S, H, CNOT\} + T$~\cite{Bravyi_2002_Fermionic_QC}. While realizing the multi-qubit Clifford gates $\{S,H,CNOT\}$ is possible in the classical metamaterials, the $T$ gate seems impossible.

The $CNOT$ gate is a two-qubit gate, which flips the target qubit if the control qubit is in the state $\ket{1}$. This is realized by a four-particle interaction expressed as the unitary operator $U_{1,2,3,4} = \exp(i\frac{\pi}{4}\gamma_1 \gamma_2 \gamma_3 \gamma_4)$~\cite{Bravyi_2002_Fermionic_QC}. Since each MZM is an eigenmode of the classical Kitaev chain, this requires implementing a non-linear interaction between different vibrational modes, which may be realized by using solitary waves. We leave this as an open problem for future study. By supplementing the single-qubit Clifford gates admissible by braiding with the $CNOT$ gate, it is possible to realize the multi-qubit Clifford gate set $\{H,S,CNOT\}$.

However, the magic $T$ gate is not achievable by the classical system because its realization requires the fusion of MZMs~\cite{Sato_2016_Majorana, Freedman_2006_FQHE}. The fusion of MZMs is realized by bringing them to a single site and measuring their total charge. As discussed in Appendix~\ref{appendix:fusion_splitting_MZMs}, if we fuse the classical MZMs, the system's topology is changed, causing a loss of information stored in MZMs (see Movie S6 for more details). Therefore, we can conclude that a universal gate set is not achievable by manipulating MZMs in the classical system.

The conclusion in this section is consistent with the prediction by Gottesman-Knill Theorem~\cite{Gottesman-Knill_theorem}, but we interpret the classical simulability in a more concrete way specific to our system. Therefore, we expect that our classical metamaterial system has achieved a highly faithful simulation of a subset of the quantum computation process attainable by the quantum Kitaev chain.

\section{Summary and Outlook}

By designing a classical mechanical metamaterial system to realize the one-dimensional Kitaev chain, we have presented a novel approach for realizing non-Abelian braiding of arbitrary pairs of classical MZMs. Our dynamical simulation demonstrates that the MZMs precisely obey the braiding statistics predicted by quantum theory, and the Yang-Baxter equations are satisfied. By properly encoding the vibrational states into a classical analog of qubit, we faithfully realize the topological computation admissible by braiding Majorana modes. Leveraging the inherent noise resilience of the Kitaev chain, our system is able to support complex classical topological computations that remain stable despite the presence of system defects.

Our study bridges two different fields, topological quantum computation and classical photonic metamaterials.
On one hand, the braiding protocols for arbitrary pairs of classical MZMs can bring new insights for the experimental realization of quantum computation based on Majorana modes. On the other hand, the topological properties predicted by quantum theory provides better mechanical stability of our system. This interaction is beneficial for both fields. Our work also demonstrates the inherent limitations of studying quantum systems by classical phononic crystals.
The lack of quantum coherence makes the faithful simulation of universal quantum computation impossible. 
Therefore, future research efforts may focus on the classically simulable parts, like the realization of the $CNOT$ gate in the current system. 
Given this intrinsic difference between classical and quantum systems, our work still shows that classical systems are useful platforms and benchmarking tools for understanding quantum phenomena in classically simulable regimes.

\begin{acknowledgements}
We thank Katia Bertoldi, Jian Li, and Pai Wang for fruitful discussions. L.C. thanks Arthur Jaffe for his support during the completion of this project.
L. C. is supported by the Army Research Office Grant W911NF-19-1-0302, and by the Army Research
Office MURI Grant W911NF-20-1-0082.
\end{acknowledgements}

\begin{appendix}

\section{Details of the dynamical matrix}\label{appendix:details_dyn_matrix}

In this section, we provide additional details of the dynamical matrix $D$ as described in \eqref{eqn:real_space_dynamical_matrix} of Section~\ref{sec:1D_Kitaev_Chain}.
The matrix $D$ can be decomposed into the onsite block and nearest neighbor coupling block as $D = D_\text{on} + D_\text{NN}$, where we leave out the trivial $\omega_0^2 I$ part which can be tuned by extra grounding springs.
Then, the onsite block $D_\text{on}$ in the $i$-th unit cell is given by:
\begin{equation} \label{eqn:D_onsite_block}
    D_\text{on} =
    \begin{array}{lc}
    \mbox{}&
    \begin{array}{cccc}u_{i}^{1}&u_{i}^{2}&u_{i}^{3}&u_{i}^{4} \end{array}\\
      \begin{array}{c}u_{i}^{1}\\u_{i}^{2}\\u_{i}^{3}\\u_{i}^{4}\end{array}&
    \begin{pmatrix}  &  & & \textcolor{RoyalBlue}{-\mu}\\  &  &\textcolor{RoyalBlue}{-\mu} &  \\  &\textcolor{RoyalBlue}{\mu} &  &  \\\textcolor{RoyalBlue}{\mu} &  &  &  \end{pmatrix}
    \end{array}
\end{equation}
where the row and column labels specify the DOFs coupled by the matrix. Similarly, the nearest neighbor coupling block $D_\text{NN}$ between the $i$-th and $(i+1)$-th cells is: 
\begin{flalign} \label{eqn:D_NN_block}
    D_\text{NN}=\begin{array}{@{}r@{}c@{}c@{}c@{}c@{}l@{}}
   & u_{i}^{1} & u_{i}^{2} & u_{i}^{3} & u_{i}^{4}  \\
   \left.\begin{array}
    {c} u_{i+1}^{1} \\u_{i+1}^{2} \\u_{i+1}^{3} \\u_{i+1}^{4} \end{array}\right(
                   & \begin{array}{c}  \\  \\ \textcolor{LimeGreen}{\frac{\Delta_{y}}{2}}\\ \textcolor{red}{\frac{t}{2}+\frac{\Delta_{x}}{2}} \end{array}
                   & \begin{array}{c}  \\  \\ \textcolor{Orchid}{-\frac{t}{2}+\frac{\Delta_{x}}{2}} \\ \textcolor{LimeGreen}{-\frac{\Delta_{y}}{2}} \end{array}
                        & \begin{array}{c} \textcolor{LimeGreen}{-\frac{\Delta_{y}}{2}} \\ \textcolor{red}{-\frac{t}{2}-\frac{\Delta_{x}}{2}} \\ \\  \end{array}
                        & \begin{array}{c} \textcolor{Orchid}{\frac{t}{2}-\frac{\Delta_{x}}{2}} \\ \textcolor{LimeGreen}{\frac{\Delta_{y}}{2}} \\  \\  \end{array}
                         & \left)\begin{array}{c} \\ \\ \\ \\ \end{array}\right.
  \end{array}
\end{flalign}
By combining the two blocks and Hermiticity, one can write down the whole dynamical matrix $D$. In \eqref{eqn:D_onsite_block} and \eqref{eqn:D_NN_block}, for clarity, we color each matrix element according to their physical realization by tunable springs, as described in Section~\ref{sec:1D_Kitaev_Chain} and Fig.~\ref{fig1}.

\section{More on the Kitaev model}
\label{appendix:review_Kitaev_chain}

\begin{figure}[t]
    \begin{center}
        \includegraphics[width=1.02\columnwidth]{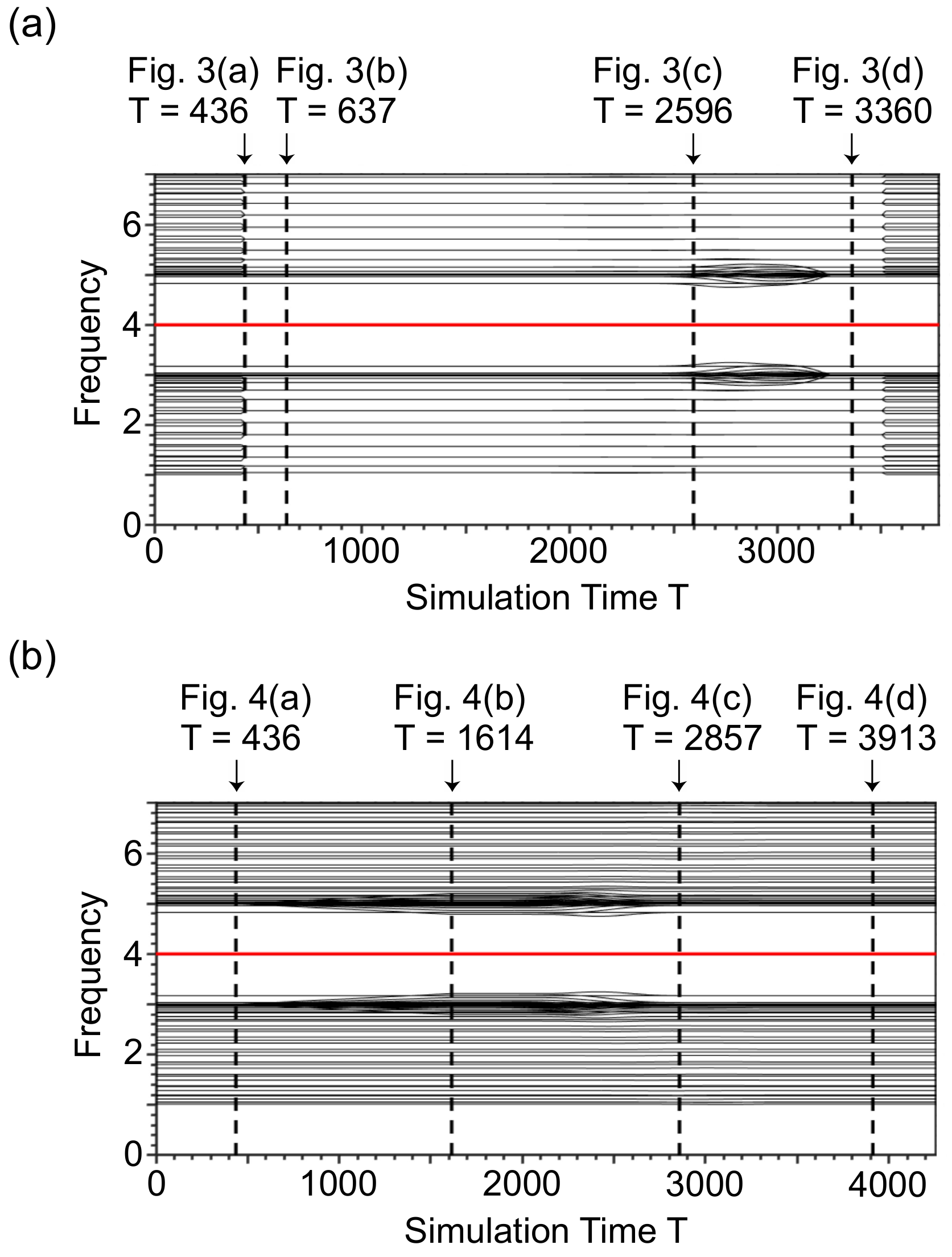}
        \caption{\textbf{The spectra of the dynamical matrix $D$ during the braiding operations} for (a) $U_{1,2}$ and (b) $U_{2,3}$. Each vertical dashed line corresponds to the time step indicated by its label. In the spectra, the bulk modes and MZMs are denoted by black and red lines, respectively.}
        \label{fig_U12_U23_spectrum}
        \vspace{-15pt}
    \end{center}
\end{figure}

We provide additional details on the one-dimensional Kitaev topological superconductor, as defined in \eqref{eqn:Kitaev_chain_Hamiltonian}. The Hamiltonian can be diagonalized by squaring it and utilizing the algebra of Pauli matrices $\sigma_{i}\sigma_{j} = \delta_{ij}I+i\epsilon_{ijk}\sigma_{k}$, to obtain the energy spectrum at a given momentum $k$:
\begin{equation}
    E(k) = \pm \sqrt{\left(\mu-t\cos(k)\right)^2+|\Delta|^2 \sin(k)^2}\;.
\end{equation}
Since the system is one-dimensional, the momentum is a scalar quantity restricted to $k\in[-\pi,\pi]$. As detailed in Kitaev's original work\cite{Kitaev_2001}, the Hamiltonian \eqref{eqn:Kitaev_chain_Hamiltonian} exhibits particle-hole symmetry, ensuring that eigenstates with energies $E(k)$ and $-E(-k)$ always appear in conjugate pairs. 
Furthermore, this symmetry enforces the presence of zero-energy eigenstates, known as Majorana zero modes (MZMs), in the topologically nontrivial phase ($|\mu|<|t|$) under open boundary conditions, where they localize at the system’s edges.

\begin{figure*}[!tbh]
    \begin{center}
    \includegraphics[width=1.8\columnwidth]{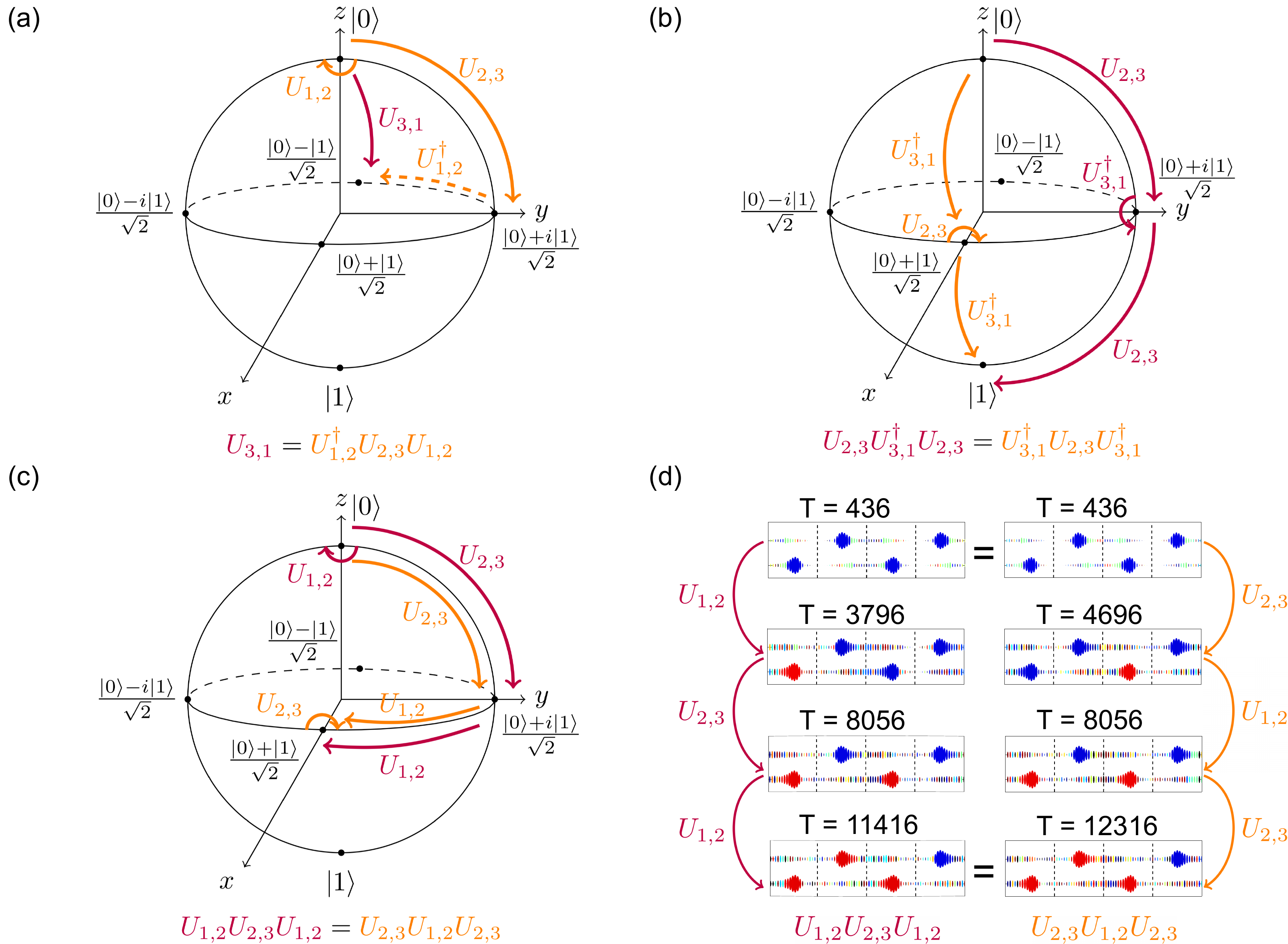}
    \caption{\textbf{The verification of braiding algebra for}: (a) $U_{3,1}$; (b) and (c) Yang-Baxter equations; and (d) MZM configurations in the verification of (c), where snapshots after each exchange operation are presented.}
    \label{fig_Yang_Bax}
    \end{center}
\end{figure*}

The bulk Hamiltonian exhibits an energy gap given by $$\Delta E = 2\sqrt{(\mu-t\cos(k))^2+|\Delta|^2 \sin(k)^2}$$ which closes and reopens at $k=0$ or $k=\pi$ when the system undergoes a transition between the topologically nontrivial ($|\mu|<|t|$) and trivial ($|\mu|>|t|$) phases. Consequently, as long as the gap remains open, the MZMs remain isolated from bulk modes. In the presence of perturbations, the integrity of the information stored in MZMs is preserved as long as the perturbation strength does not induce a gap-closing transition. Thus, the spectral gap serves as the fundamental mechanism for topological protection.

In mechanical metamaterials, the frequency spectrum serves as the classical analog of the energy spectrum in quantum systems. 
Consequently, we expect that the spectral gap in the classical Kitaev chain can similarly provide topological protection. 
This requires that the gap of the dynamical matrix $D$ in \eqref{eqn:real_space_dynamical_matrix} remains open throughout the braiding process. 
This condition is explicitly verified in Fig.~\ref{fig_U12_U23_spectrum}, where subfigures (a) and (b) illustrate the frequency spectra of $D$ during the $U_{1,2}$ and $U_{2,3}$ braiding operations, as depicted in Figs.~\ref{fig_U12} and \ref{fig_U23}, respectively. 
Each vertical dashed line represents a time step in the braiding sequence, as indicated by the labels. 
In these spectra, bulk mode frequencies are shown in black, while the MZM frequencies are highlighted in red. Notably, the spectral gaps remain open throughout the braiding processes, and the MZMs are always isolated. This ensures the fault-tolerant computation discussed in Section~\ref{sec:FT_braiding}.

\section{Braiding and Yang-Baxter Equations} \label{appendix:Yang_Baxter_Eqns}

In this section, we provide additional verifications on the braiding algebra in the classical Kitaev chain, as illustrated in Fig.~\ref{fig_Yang_Bax}. 
First, as stated in the main text, the exchange of the first and third MZMs can be implemented via the relation $U_{3,1} = U_{1,2}^\dagger U_{2,3} U_{1,2}$ (see Section~\ref{sec:single_qubit_Clifford}), which is explicitly confirmed in Fig~\ref{fig_Yang_Bax}(a). 
Second, we verify the Yang-Baxter equations, as shown in Figs.~\ref{fig_Yang_Bax}(b), (c), and (d). Before presenting the detailed verification, we briefly review these equations.

Verifying the braiding of $n$ classical MZMs and ensuring that they satisfy the Yang-Baxter equations is essential to establishing that the system realizes a representation of the braid group on $n$ strands. For completeness, we reproduce these equations here:
\begin{align} \label{eqn:YB_Eqn_Appendix}
    U_{i,i+1} U_{j,j+1} &= U_{j,j+1} U_{i,i+1}, \quad |i-j|>1\;, \notag\\
    U_{i,i+1} U_{j,j+1} U_{i,i+1} &= U_{j,j+1} U_{i,i+1} U_{j,j+1}, \quad |i-j| = 1\;,
\end{align}
where the first equation expresses that braiding operations involving well-separated MZMs commute, a property that holds trivially in our system. The second equation encapsulates a nontrivial equivalence between two distinct sequences of neighboring MZM exchanges, which we explicitly verify in the remainder of this section.

\begin{figure*}[!tbh]
    \begin{center} 
    \includegraphics[width=1.8
    \columnwidth]{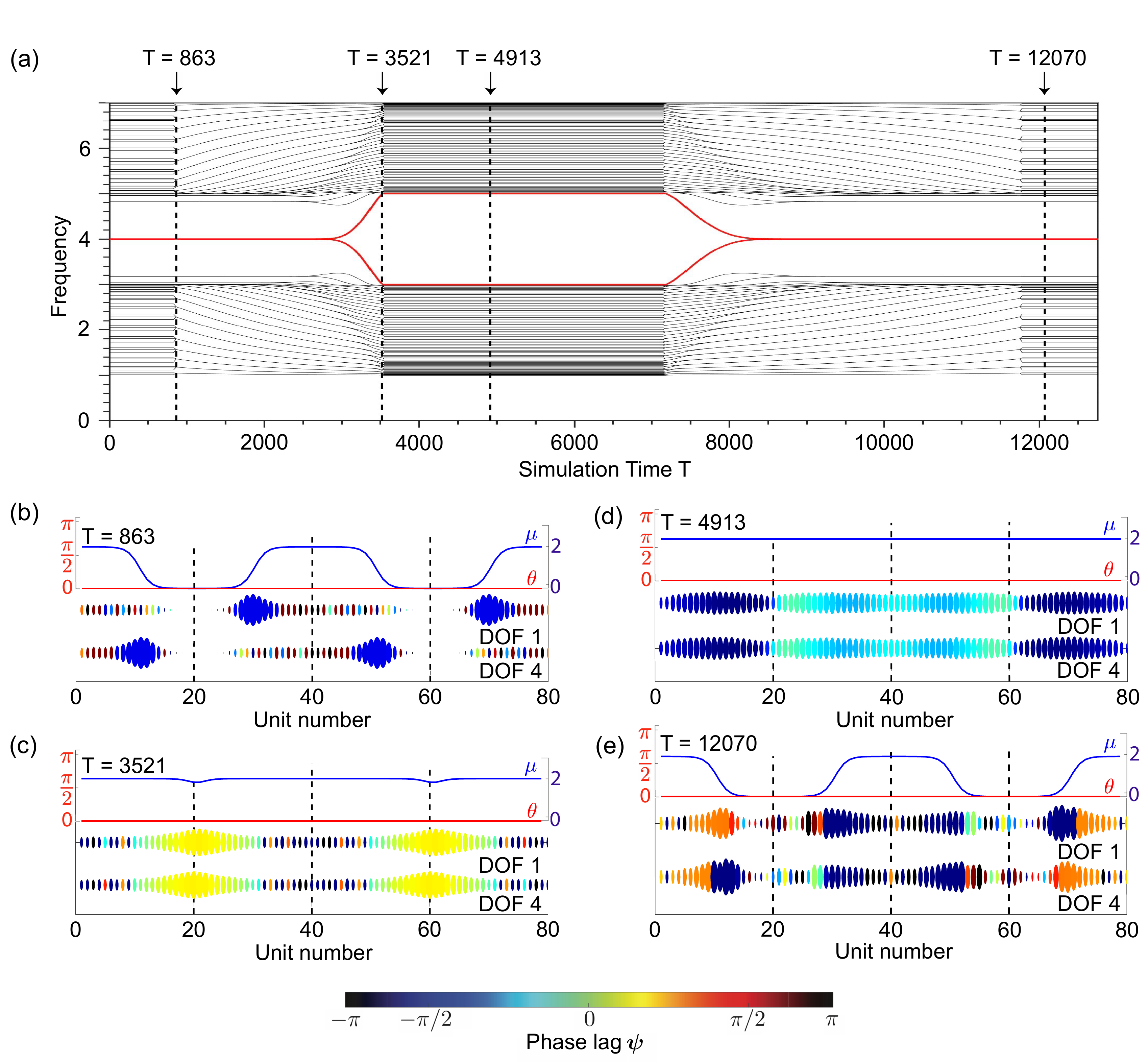}
        \caption{\textbf{The Fusion and Splitting of Majorana Zero Modes (MZMs).} (a) The frequency spectrum of the process, where the dashed lines correspond to the representative steps in (b)-(d). (b) The initialization of MZMs. (c) The fusion of MZMs by bringing each pair of domain walls to the same position. (d) After fusion, the MZMs completely disappear. (e) The regeneration of MZMs by separating the domain walls.}
    \label{fig_fusion_spectrum}
    \end{center}
\end{figure*}

Given four MZMs $\gamma_1,\gamma_2,\gamma_3$ and $\gamma_4$, the second Yang-Baxter equation can be expressed in two equivalent forms: 
\begin{equation}
    U_{2,3}U_{3,1}^\dagger U_{2,3} = U_{3,1}^\dagger U_{2,3}U_{3,1}^\dagger
\end{equation}
and 
\begin{equation} \label{eqn:Yang_Baxter_U_12_U_23}
    U_{1,2}U_{2,3}U_{1,2} = U_{2,3}U_{1,2}U_{2,3}
\end{equation}
as illustrated in Fig.~\ref{fig_Yang_Bax}(b) and Fig.~\ref{fig_Yang_Bax}(c), respectively. While both equations have been verified, for simplicity, we present only the verification of the latter. As shown in Fig.~\ref{fig_Yang_Bax}(d), the left and right columns depict the evolution of the system starting from the initial state $\ket{0}$ under the sequence of braiding operations prescribed by the left-hand side and right-hand side of \eqref{eqn:Yang_Baxter_U_12_U_23}, respectively. The final configurations are identical in both cases, confirming the validity of the Yang-Baxter equation in this instance. Furthermore, we have verified that the equation holds for any arbitrary initial configuration of MZMs, establishing that the braiding operations form a representation of the Artin braid group.

\section{The fusion and splitting of MZMs} \label{appendix:fusion_splitting_MZMs}

As discussed in Section~\ref{sec:UQC}, fusion is necessary for realizing universal quantum computation using MZMs. 
In this section, we demonstrate that fusing MZMs in the classical Kitaev chain will destroy the encoded logical information, compromising the fault tolerance.

We simulate the fusion and splitting processes of two pairs of classical MZMs, as shown in Fig.~\ref{fig_fusion_spectrum}. The frequency spectrum is shown in Fig.~\ref{fig_fusion_spectrum}(a), where the four vertical dashed lines correspond to the four representative steps during the simulation. In the first step, we initialize four MZMs at the domain walls (see Fig.~\ref{fig_fusion_spectrum}(b)). Then, we fuse each pair of MZMs by moving the domain walls close to each other, as shown in Fig.~\ref{fig_fusion_spectrum}(c). After the system equilibrates, the MZMs completely disappear (see Fig.~\ref{fig_fusion_spectrum}(d)). Then, we try to regenerate these MZMs by splitting the domain walls, as illustrated in Fig.~\ref{fig_fusion_spectrum}(e). However, the shape and phase information in the final configuration differs significantly from the initial one, indicating a loss of logical information caused by the fusion.

This information loss is a direct consequence of the topological phase transition, as indicated by the spectrum in Fig.~\ref{fig_fusion_spectrum}(a). When we implement the fusion, the MZMs are no longer isolated from the bulk modes. Thus, the topological protection from the energy gap no longer exists. The system becomes vulnerable to perturbations, which results in information loss. From another perspective, bringing domain walls to the same positions effectively causes the whole one-dimensional chain to be topologically trivial (see Figs.~\ref{fig_fusion_spectrum}(c)(d)), so we cannot expect the information in MZMs to be preserved.

Our results are also insightful for topological quantum computation based on MZMs, where fusion is necessary for achieving universal computational power. When fusing MZMs in a topological superconductor, the loss of logical information may be unavoidable. Simultaneously, the topological phase transition will invalidate the topological protection. Thus, it may be both interesting and meaningful to figure out a noise-resilient method for realizing the universal gate set.

\end{appendix}

\bibliography{References}

\end{document}